\documentclass[twocolumn,amsmath,superscriptaddress,amssymb,amsfonts,prl,floatfix]{revtex4}

\usepackage{graphicx, color,rotating} 
\usepackage{dcolumn} 
\usepackage{bm} 
\usepackage{epsfig}
\usepackage{bbm}

\begin{document}

\title{\huge Electronic dispersion anomalies
in the iron pnictide superconductor Ba$_{1-x}$K$_x$Fe$_2$As$_2$}

\author{\bf Andreas~Heimes}
\affiliation{Institut f\"ur Theoretische Festk\"orperphysik, Karlsruhe Institute of Technology, D-76131 Karlsruhe, Germany}

\author{\bf Roland~Grein}
\affiliation{Institut f\"ur Theoretische Festk\"orperphysik, Karlsruhe Institute of Technology, D-76131 Karlsruhe, Germany}
\affiliation{DFG-Center for Functional Nanostructures, Karlsruhe Institute of Technology, D-76128 Karlsruhe, Germany}

\author{\bf Matthias~Eschrig}
\affiliation{Institut f\"ur Theoretische Festk\"orperphysik, Karlsruhe Institute of Technology, D-76131 Karlsruhe, Germany}
\affiliation{DFG-Center for Functional Nanostructures, Karlsruhe Institute of Technology, D-76128 Karlsruhe, Germany}
\affiliation{Fachbereich Physik, Universit\"at Konstanz, D-78457 Konstanz, Germany}

\begin{abstract}
{\bf
The pairing mechanism in the iron-pnictide superconductors is still
unknown. However, similarities to the cuprate high-temperature superconductors suggest that a similar mechanism may be at work.
Recently, careful experimental studies of the spin excitation spectrum
revealed, like in the cuprates, a strong temperature dependence in the normal
state and a resonance feature in the superconducting state.
Motivated by these findings, we develop a model of electrons interacting
with a temperature dependent magnetic excitation spectrum based on these experimental observations.
We apply it to analyse angle resolved photoemission and tunnelling
spectra in Ba$_{1-x}$K$_x$Fe$_2$As$_2$.
We reproduce in quantitative agreement with experiment
a renormalisation of the quasiparticle dispersion
both in the normal and the superconducting state,
and the dependence of the quasiparticle linewidth on binding energy.
We estimate the strength of the coupling between electronic and spin
excitations. Our findings support the possibility of a
pairing mechanism based dominantly on such a coupling.
}
\end{abstract}
\maketitle

Shortly after the discovery of high-temperature superconductivity in Fe-based pnictide compounds \cite{Hosono2008} a magnetic Cooper-pairing mechanism was proposed \cite{Mazin08}, while electron-phonon interaction as primary pairing mechanism was found to be  unlikely \cite{Boeri08}. This conjecture is supported by the proximity of antiferromagnetism and superconductivity in the phase diagram \cite{Zhao2008,Luetkens2008,Chu2009,Nandi2010,Paglione2010}. 
The spin excitation spectrum in pnictides shows pronounced similarities
with other superconductors where a magnetic pairing mechanism is under debate.
In particular, the strong temperature dependence of the normal state spin 
excitations studied recently by Inosov {\it et al.} \cite{Inosov2009}, as well as
the presence of a spin resonance feature in the superconducting state 
\cite{Christianson2008} are prominent features also present in cuprates,
as well as in some heavy fermion superconductors. 

Furthermore, angle resolved photoemission (ARPES) measurements reveal a sharp Fermi surface consisting of electron-like and hole-like pockets that exhibit comparable superconducting order parameter amplitudes and are nearly nested by the antiferromagnetic wave vector $\bm Q$ \cite{Ding2008,Nakayama2009}, a reason to believe that magnetic and electronic order are closely connected \cite{MazinSchmalian2009}. This raises the question how strongly electrons couple to spin fluctuations, as those lead to an effective electron-electron attraction, if the order parameter changes the sign on the different pockets \cite{Parker2008,Chubukov2008,Maier2008}. Investigating the low energy dispersion anomalies, whose position and shape can be traced back to a coupling of bosonic modes, has proven to be a powerful method for obtaining this information \cite{Richard2008, Hasan2008, Eschrig2006}.
Recent experimental studies on the superconducting state spectral function of Ba$_{1-x}$K$_x$Fe$_2$As$_2$ measured by ARPES find such an anomaly, which appears as a kink in the dispersion relation at about $25 \, $meV \cite{Richard2008}.
On the other hand, inelastic neutron scattering (INS) studies of this compound reveal the development of a resonant spin excitation in the superconducting state \cite{Christianson2008}. The temperature dependence of both neutron intensity and the self-energies extracted from ARPES follow an order parameter like evolution. This strongly indicates that the spin resonance indeed accounts for the observed anomaly.

Below, we present a theoretical model that explains the anomalous features in the dispersion relation by a coupling of electrons to a spin fluctuation spectrum 
that (a) reproduces the experimentally observed temperature dependent spectrum
in the normal state and (b) features a low energy resonance in the superconducting
state. 
We show that coupling to such a spin-fluctuation spectrum leads to a renormalisation of the quasiparticle dispersion even in the normal state and can account for the renormalisation factor necessary to match density functional theory (DFT) calculations to experiment. Furthermore, taking into account the resonance in the superconducting state, we are able to quantitatively reproduce experimental ARPES results, including low-energy and high-energy renormalisation of the dispersion and the energy dependence of the quasiparticle linewidth.
Our theoretical SIN tunnelling spectra obtained for the same parameter set are consistent with recent experimental observations that show an asymmetric lineshape of the tunnelling conductance \cite{Shan2010}.\\
\begin{figure*}[hbtp]
  \begin{center}
	\epsfig{figure=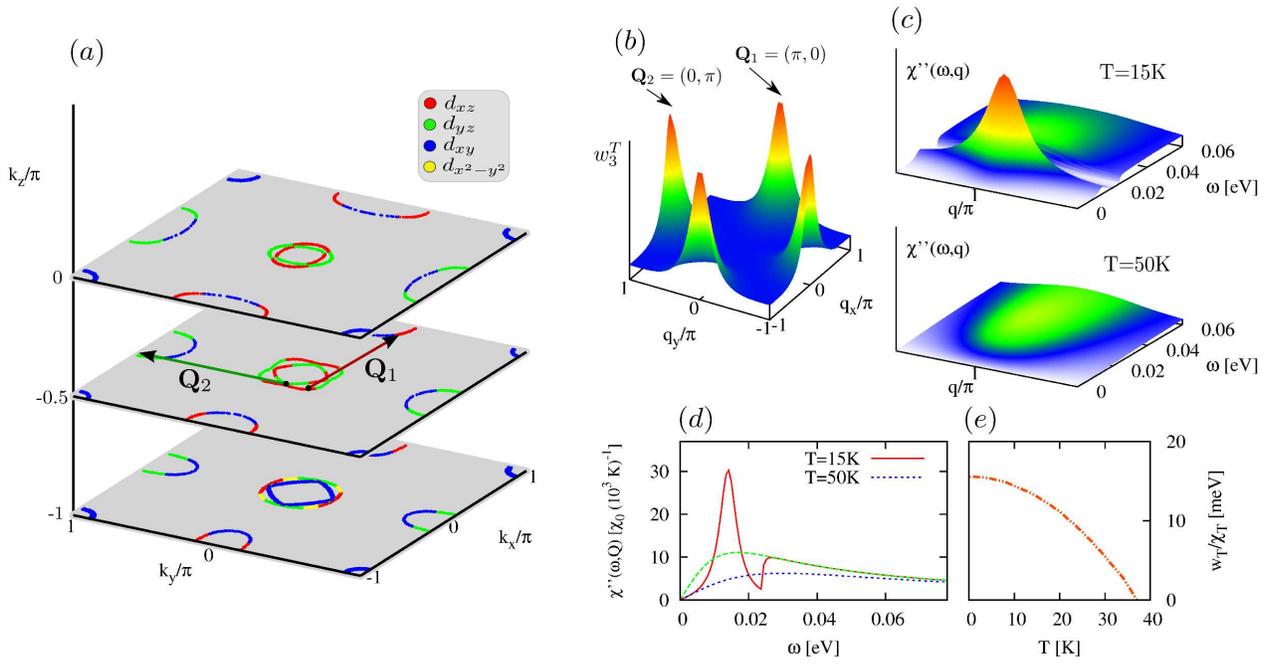, angle=270, width=0.95\textwidth}
    \caption{\textbf{Fermi surface geometry and characteristics of the
spin fluctuation mode.}
($\bm a$) The dominant orbital contributions to the bare Fermi surface at $k_z=-\pi,\,-\pi/2,\,0$. Red, green, blue and yellow correspond to the $d_{xz},\,d_{yz},\,d_{xy}, \, d_{x^2-y^2}\,$-orbitals, respectively. ($\bm b$) 
Momentum dependence of the spin fluctuation mode.
The mode is centred at wavevectors $\pm\bm Q_1=(0,\pm\pi)$ and $\pm\bm Q_2=(\pm\pi,0)$. ($\bm c$) Energy-momentum dependence of the spin susceptibility in the normal state at $T=50\,K$ (bottom) and in the superconducting state at $T=15\,K$ (top). ($\bm d$) The shape of the spectral function $\chi''(\omega,\bm Q)$ is shown for $T=50\,K$ (blue line) and $T=15\,K$ (red line). The dashed green line shows the behaviour for $T=$15 K
if the material were in the normal state.
($\bm e$) The ratio of the resonance and the continuum spectral weight, $w_T/\chi_T$ as function of temperature.  Concerning temperature dependence of $\chi''(\omega, \bm q)$ see Fig. \ref{S1}.}
    \label{fig1}
  \end{center}
\end{figure*}

\noindent {\bf Spin fluctuation spectrum and resonance mode}\\
\noindent
The appearance of a resonance in the dynamic magnetic susceptibility 
upon entering the superconducting state is well known from cuprate superconductivity. 
It is situated at an energy below the particle-hole continuum
and is peaked around the antiferromagnetic wavevector.
Although there are various theoretical models for the spin-spin response function,
here we prefer to not rely on a specific theoretical model,
but rather use a semi-phenomenological approach where the
spin susceptibility is taken from experiment. This approach was very successful 
in the case of cuprates \cite{Eschrig2006}, and has the advantage of 
being independent of specific theoretical assumptions and approximations.
In addition, most of the magnetic correlations 
are included automatically within this approach by relying on experimental results.

A detailed investigation of the spin dynamics in the normal state and the temperature dependence of the spin resonance in pnictides was recently performed on optimally doped BaFe$_{1.85}$Co$_{0.15}$As$_2$ by Inosov {\it et al.} \cite{Inosov2009}.  
They observe that the imaginary part of the normal state susceptibility 
near the antiferromagnetic wave vectors $\bm Q_\alpha $
obeys in good approximation the Ornstein-Zernicke form
\begin{eqnarray}
\label{eqn1}
\chi_{c,n}(\omega,\bm q)= \sum_{\alpha =1,2}
	 \frac{b_{n,\alpha }\,\chi_T}{1+\xi_T^2\,\left|\bm q - \bm Q_\alpha \right|^2-\imath (\omega/\Omega_{max}^T)},
\end{eqnarray}
where $\chi_T=\chi_0/(T+\theta)$ controls the strength of magnetic correlations, 
$\Omega^T_{max}=\Omega_0(T+\theta)$ sets the typical spin-fluctuation energy scale,
$\xi_T=\xi_0/\sqrt{T+\theta}$ is the magnetic correlation length and $\theta$ is the Curie-Weiss temperature.  

With decreasing temperature, spectral weight is shifted towards lower energies. Below $T_c$ the spectrum becomes gapped and a
resonance appears at an energy that follows approximately an order parameter like evolution, i.e. $\Omega_{res}^T=\Omega_r\, \sqrt{1-T/T_c}$.
In order to model the resonance below the quasiparticle continuum we assume a (sharply peaked) Lorentzian in energy and a separable form for the momentum dependence,
\begin{equation}
\label{eqn2}
\chi_{r,n}(\omega,\bm q) =
\frac{w^T_n (\bm q)}{\pi}
\frac{2 \Omega_{res}^T }{(\Omega_{res}^T)^2-(\omega +\imath \Gamma_{res})^2}
\quad
\end{equation}
with the weight functions
\begin{eqnarray}
\label{eqn2a}
w^T_n ({\bm q})&=& 
\sum_{\alpha =1,2}
\frac{b_{n,\alpha } \, w_T}{1+\xi_r^2\,\left| {\bm q} -{\bm Q}_\alpha \right|^2} .
\end{eqnarray}
In order to take care of the periodicity in reciprocal space we replace
the factors $|\bm q -\bm Q_\alpha |^2$ in Eqs.~\eqref{eqn1} and \eqref{eqn2a}
by $4[ \sin^2(\frac{q_x-Q_{\alpha x}}{2})  + \sin^2(\frac{q_y-Q_{\alpha y}}{2} )]$, 
and we neglect the dependence on $q_z$ for our purpose 
as it varies weakly \cite{Park_et_al_2010,Lumsden2009}.
The resulting momentum dependence at the resonance energy is shown in Fig.~\ref{fig1}(b), with the two antiferromagnetic wavevectors ${\bm Q}_1=(\pi,0)$ and ${\bm Q}_2=(0,\pi)$. These wavevectors connect Fermi surface sheets with same orbital character ($n=1,2 \cdots 5$ corresponding to the five Fe orbitals $d_{xz}$, $d_{yz}$, $d_{xy}$, $d_{x^2-y^2}$, $d_{3z^2-r^2}$), as illustrated in Fig.~\ref{fig1}(a). In our model we consider only the main orbital contributions to quasiparticle scattering resulting from the $n=1,2,3$ orbitals and describe the different momentum variations 
with the help of the parameter $b_{n,\alpha }$ ($=1/2$ for $n\alpha =11,22,31,32$ and $=0$ else).
 
For the gapped continuum (in the absence of detailed experimental 
data) we make the approximation that 
for $\omega $ above a temperature dependent continuum threshold, 
$\omega>2 \bar \Delta (T)$,
it is identical to its normal state value, Eq.~\eqref{eqn1},
at the respective temperature. We note that this approximation is 
excellent except possibly close to the continuum threshold.
Here, $\bar \Delta$ is the mean superconducting gap at the nested Fermi surfaces. 
Thus, denoting the imaginary part $\mbox{Im} \chi$ of the retarded dynamical susceptibility with $\chi''$,
we approximate the superconducting state spectrum of the susceptibility by
\begin{equation}
 \chi''_{sc,n}(\omega ,{\bm q} )=
\left\{
\begin{array}{ccc}
\chi''_{r,n} (\omega ,{\bm q})& \mbox{for} & |\omega|< 2\bar \Delta\\
\chi''_{c,n} (\omega ,{\bm q})& \mbox{for} & |\omega|\ge 2\bar \Delta\\
\end{array}
\right. ,
\end{equation}
and calculate the real part by exploiting Kramers-Kronig relations.
In order to reduce the number of parameters we employ a
local sum rule to determine the ratio between resonance and continuum part,
$w_T/\chi_T$.
It is chosen so that the energy and momentum integrated spin structure factor,
$\int_{-\infty}^{\infty } d\omega \int d{\bm q} S(\bm q,\omega )$, with
$S(\bm q, \omega)=2\, \hbar \, \chi''(\bm q,\omega) \,/(1-e^{-\hbar \omega/k_B T})$ remains temperature independent in the normal as well as the superconducting phase (for technical details see the Methods section). 
The overall weight $\chi_0$ will be combined with the coupling constant $g$
below, and it is only the combined quantity $g^2\chi_0$ that is of relevance for
our results. However, as discussed below,
from experimental values of $\chi_0$ and from a determination of 
the fit parameter $g^2\chi_0$ an estimate of the coupling constant 
can be obtained.  

\begin{table}
\caption{\label{tab1} Parameter Set used for Ba$_{0.6}$K$_{0.4}$Fe$_2$As$_2$
}
\begin{ruledtabular}
\begin{tabular}{ccccccc}
$\frac{w_{T= 0K}}{\chi_{T= 0K}}$ &
$\Omega_0$ &
$\xi_0$ &
$\theta$ &
$\xi_r$ &
$\Omega_r$ &
$\Gamma_{res}$\\
15.6 meV &
0.375 $\frac{\text{meV}}{\text{K}}$   &
1.07 K$^{1/2}$&
30 K &
2 &
15.5 meV &
3 meV\\
\end{tabular}
\end{ruledtabular}
\end{table}

We apply this model to hole-doped Ba$_{0.6}$K$_{0.4}$Fe$_2$As$_2$, where
data are available both for spin excitations as well as for electronic excitations,
and where also the most detailed experimental angle resolved photoemission data exist.
In FIG.~\ref{fig1} the momentum and energy dependence for the parameter set in TABLE~\ref{tab1} is presented. The resonance in the spin excitation appears at an energy $\Omega_{res}^{T=7K}\approx 14\, $meV below the quasiparticle continuum $\omega < 2 \bar \Delta(15$K$) \approx 24 \, $meV \cite{Christianson2008}. In momentum space the mode is peaked around the wave vectors $\bm Q_1=(\pi,0,q_z)$ and $\bm Q_2=(0,\pi,q_z)$ with correlation length $\xi_r$ of nearly twice the lattice constant [FIG.~\ref{fig1}(b)]. The momentum dependence in $z$-direction is assumed to vary weakly as motivated by \cite{Park_et_al_2010,Lumsden2009}. Comparing the energy-momentum distribution in the normal and the superconducting state [FIG.~\ref{fig1}(c)], we see that 
low-energy spectral weight is shifted into the resonance when the continuum 
gap opens below $2\bar \Delta$. In addition, the sharpened momentum distribution leads to an enhancement of the resonance spectral weight. In FIG.~\ref{fig1}(e) we show the numerically calculated ratio $w_T/\chi_T$, which features this evolution and is in excellent agreement with the functional form $w_T/\chi_T=w_{T=0K}/\chi_{T=0K}\,(1-T^2/T_c^2)$. As we will see later on, such a modelling will lead to an effect on the electronic dispersion that fits well with experimental observations.\\

\noindent {\bf Coupling to spin fluctuations}\\
\noindent
We are interested in the renormalisation of the fermionic dispersion as a result of the coupling of electrons to the spin fluctuation mode. The idea is to extract the influence of the resonance by comparing the superconducting and normal state dispersion from which self-energy effects are directly inferred. This allows for an immediate comparison with experiment.
Following the approach for the cuprates \cite{Eschrig2006}, one could assume that the momentum dependence in Eq.~({\ref{eqn1}) and Eq.~({\ref{eqn2}) exclusively chooses the Fermi surface sheets that are coupled by the mode. However, fluctuation exchange (FLEX) approaches have shown that the magnetic mode predominantly scatters between states with the same orbital character \cite{Kemper2010, Stanescu2008}.

To take this into account, we employ a tight-binding fit in orbital basis which was obtained from the DFT band structure of BaFe$_2$As$_2$ by Graser {\it et al.} \cite{Graser2010}. Details are discussed in the Methods section.

\begin{figure*}[hbtp]
  \begin{center}
     \epsfig{figure=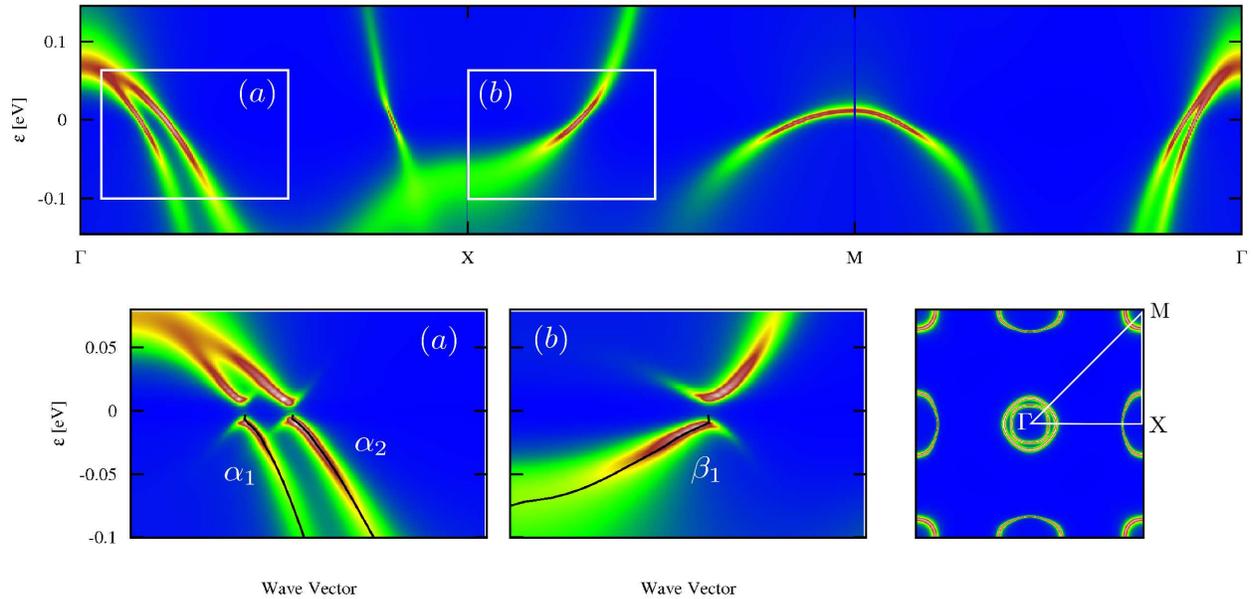, angle=270, width=0.95\textwidth}
    \caption{\textbf{Electronic spectral intensity.} On top: spectral function in the normal state for $T=50\,K$ along a cut in the 1st Brillouin zone for $k_z=0$. At bottom: the same for the superconducting state at $T=15\,K$ for the square regions
indicated (a) and (b) in the top panel. The black lines show
MDC-derived dispersions for the ($\bm a$) $\alpha_1$ and $\alpha_2$ band and the ($\bm b$) $\beta_1$ band. At the lower right the renormalised Fermi surfaces including the group notations are shown.}
    \label{fig2}
  \end{center}
\end{figure*}
In FIG.~\ref{fig1}(a) we show the bare Fermi surface for $k_z=-\pi,\,-\pi/2$ and $0$, corresponding to a bc tetragonal unit cell and the I4/mmm symmetry of the crystal.  
The hole pockets at $(0,0,k_z)$ and $(\pi,\pi,k_z)$ are nearly nested to the electron pockets at $(0,\pi,k_z)$ and $(\pi,0,k_z)$ by the wave vectors $\bm Q_1$ and $\bm Q_2$, even when taking into account the orbital characters \cite{Kemper2010, Stanescu2008, Graser2009, Graser2010}.
We couple electrons to the spin fluctuation spectrum with a coupling constant $g$, which is assumed to be independent of energy, momentum and orbital number.

In ARPES experiments the intensity of photo electrons is proportional to $f(\epsilon)A(\epsilon,\bm k)$, where $f$ is the Fermi distribution function. Dispersions obtained from those usually differ from DFT calculations by an renormalisation factor of $2$ \cite{Ding2008_2, Lu2009, Sekiba2009}. In the upper panel of FIG.~\ref{fig2} we present the influence of a coupling to spin fluctuations in the normal state.
It shows an intensity plot of the spectral function $A(\epsilon, \bm k)$ 
along a cut in the 1$^{\rm st}$ Brillouin zone, obtained 
numerically as explained in the Methods
section.
The band maxima and minima as well as the Fermi velocity very well align with experimentally observed values \cite{Ding2008_2}. 
By choosing the coupling constant $g$ to be the same for all orbitals the band curvature becomes strongly enhanced leading to the observed shallow electron pocket at the $X$-point. One clearly sees that the bands are bent by a factor $1.5-2$ and that the quasiparticle width increases with energy. Thus coupling to spin fluctuations gives an essential contribution to the above mentioned renormalisation factor. However the Fermi surface remains nearly unaffected as can be seen by comparing the Fermi surface at 50K on the right bottom of FIG.~\ref{fig2} with the bare one in FIG.~\ref{fig1}(a).  \\

\noindent {\bf Superconducting order}\\
\noindent
The origin of the pairing instability may well be related to the spin fluctuation continuum, as demonstrated by a recent FLEX 
calculation \cite{Schmalian2010}. 
Our model is restricted to the low energy region in the spin excitation spectrum,
and does not precisely treat the incoherent high-energy part.
However, this part considerably contributes to pairing, whereas
the energy range of interest here
($|\omega |< \omega_c=200\,$meV) gives only a partial contribution (about 40-50\%) to the value of the superconducting order parameter.  
Thus, we add in our theory
a contribution $\Delta_{\bm k}$ to the order parameter that results from the 
incoherent high-energy part of the spin-fluctuation spectrum beyond 200 meV.

The order parameter is chosen to have an $s^\pm$-symmetry 
(here and in the following the unit of length is the in-plane lattice constant $a$)
\begin{equation}
 	\Delta_{\bm k}(T)=\Delta_0(T)\,\cos(k_x) \, \cos(k_y).
\label{eqn3}
\end{equation}
This pairing state is supported by experiment as well as numerical calculations \cite{Ding2008, Nakayama2009, Thomale2009, Graser2009, Graser2010} .
The magnitude of the superconducting gap
was observed to be $\bar \Delta($15K$) \approx 12 \, $meV at the inner hole-like pocket as well as the corresponding nested pockets \cite{Ding2008,Nakayama2009}. We choose $\Delta_0$ in Eq.~(\ref{eqn3}) so that the renormalised gap reaches the experimentally observed value at these particular points in the Brillouin zone, which gives
$\Delta_0(0)=18.1$ meV. The renormalised gap is obtained by taking into account
a renormalisation factor as discussed in the Methods.
For the temperature dependence (in the absence of detailed data) we assume
the form $\Delta_0(T)=\Delta_0(0)\sqrt{1-T/T_c}$.\\

\noindent {\bf Dispersion and linewidth: self-energy effects}\\
\noindent
In the superconducting state the dispersion features in FIG.~\ref{fig2}(a) are
modified due to (a) the appearance of the superconducting gap, and (b)
the modifications in the spin excitation spectrum that is coupled to the
conduction electrons.
In the lower panel of FIG.~\ref{fig2} extracts from the hole-like and electron-like pockets at the $\Gamma$- and $X$-point are presented for 15 K, well in the superconducting state. At first glance the intensity plots show no clear hint of the bosonic resonance, in contrast to the eye-catching break features in the cuprates \cite{Kaminski2001}. 
In order to extract a quantitative effect we need to analyse differences
between the dispersions and linewidths in the normal and superconducting state 
in detail.

Because in the case of pnictides multiple orbitals are involved in the electronic
spectra, the linewidth function and the renormalisation of the dispersion 
cannot be related in an easy way to theoretically obtained self energies as
it has been possible in the case of the cuprates. For this reason we 
follow here a different path and extract quantities from theoretically
obtained spectral functions exactly as they are extracted in the ARPES experiments
\cite{Richard2008, Hasan2008} from measured spectra.
For fixed energy the momentum dependence of the spectral function 
(a so-called momentum distribution curve, or MDC, see inset to FIG.~\ref{fig3}) is peaked, with the peak often 
well approximated by a Lorentzian.
If the self energy does not vary much as a function of momentum over the width
of such an MDC peak, the MDC spectral function is
of the form
\begin{equation}
\label{eqn12}
A_\epsilon ({\bm k})= \frac{1}{\pi} \frac{\Sigma''_{\epsilon, k_\epsilon }}{
\{{\bm v}_\epsilon ({\bm k}-{\bm k}_\epsilon)\}^2
+ \{\Sigma''_{\epsilon, k_\epsilon }\}^2},
\end{equation}
where ${\bm k}_\epsilon $ determines the maximum point of the MDC,
and ${\bm v}_\epsilon= (\partial {\bm k}_\epsilon /\partial  \epsilon )^{-1}$
\cite{Norman01}. 
In the bottom panel of FIG.~\ref{fig2} we show dispersions 
obtained from MDC maxima of the $\alpha_{1,2}$ and the $\beta_1$ band as black
curves. 
Furthermore,
we obtain the linewidth in analogy to experiment by fitting
a Lorentzian \eqref{eqn12} to the theoretically obtained ARPES spectra.

We now discuss the thus obtained quantity $\Sigma''(\epsilon ) \equiv \Sigma''_{\epsilon, k_\epsilon }$ that determines the linewidth of the MDC.
Assuming that it weakly depends on momentum in a small region around $k_\epsilon$ it is given by $\Sigma''_{\epsilon, k} \approx {\bm v}_\epsilon \, \delta {\bm k}_\epsilon$, 
and $|\delta {\bm k}_\epsilon |$ is the full width at half maximum (FWHM) of the Lorentzian in reciprocal space in direction of ${\bm v}_\epsilon $.
We underline that
it is important to distinguish between the calculated self-energy 
$\Sigma^{R}_n(\epsilon )$ for the $n^{\rm th}$ orbital, as discussed
in the Methods, and the experimentally motivated
quantity $\Sigma''(\epsilon )$ extracted by
fitting a Lorentzian to the theoretically obtained ARPES spectra.
Since the Green's function matrix in orbital space
cannot be inverted analytically there is no simple correspondence between the two
quantities above. We will call the experimentally motivated 
quantity $\Sigma(\epsilon )$ 
imaginary part of the {\it effective} self energy below.

\begin{figure}[t]
  \begin{center}
     \epsfig{figure=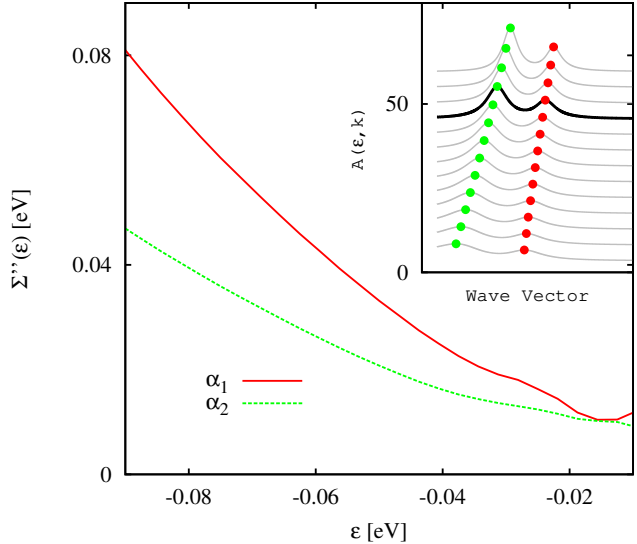, angle=270, width=0.5\textwidth}
    \caption{\textbf{Scattering rate:} Imaginary part of the effective self-energy at $T=15\,K$ determined by the width of the Lorentzian fits (centred at the
points indicated in the inset) times the group velocity at the given energy, i.e. $\Sigma''_{\epsilon}=v_\epsilon \, \delta k_\epsilon$, at the $\alpha_1$ (red) and the $\alpha_2$ (green) band.
For temperature dependent results see Fig. \ref{S2}.}
    \label{fig3}
  \end{center}
\end{figure}
\begin{figure}[h]
  \begin{center}
     \epsfig{figure=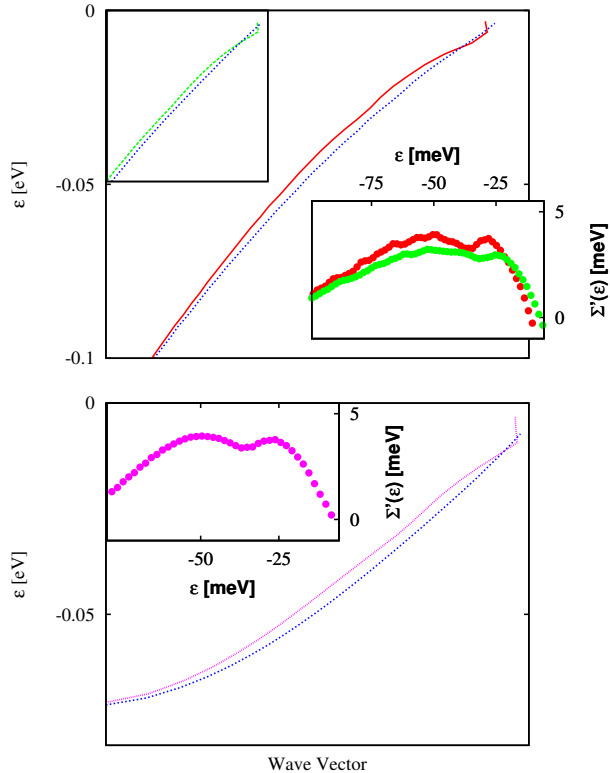, angle=270, width=0.5\textwidth}
    \caption{\textbf{MDC-derived dispersions}, obtained from the centre positions of
the Lorentzian fits as shown in the inset of Fig.~\ref{fig3},
of the $\alpha_1$ (red) and $\alpha_2$ (green) band as well as the $\beta_1$ band (magenta) in the superconducting state at $T=15\,K$. 
The blue curves are the dispersions in the normal state at $T=50\,K$ respectively. The insets show the real part of the effective self energy $\Sigma'(\epsilon)=\Sigma'_{\epsilon,k_\epsilon}$.
For temperature dependent results see Fig. \ref{S3}.}
    \label{fig4}
  \end{center}
\end{figure}
As can be seen in FIG.~\ref{fig3} the imaginary part of the effective self-energy exhibits a linear dependence for higher energies, i.e. $\Sigma''(\epsilon) \propto \epsilon$, consistent with marginal Fermi liquid theory \cite{Littlewood&Varma1991}. This results from the coupling to the continuum, in particular from the slow decay of
the spin fluctuation spectrum towards high energies.
We note that the correct magnitude of the linear in $\epsilon $ high-energy
part of $\Sigma'' (\epsilon )$ restricts the coupling strength $g^2\chi_0$.
We are able to reproduce the experimental observations \cite{Richard2008} with a 
coupling strength $g^2\chi_0=1.17 \times 10^3 \, \mu_{\rm B}^2$eV~K.
With this, all parameters of the theory are fixed by the high-energy behaviour of
the electronic spectra, and consequently
the magnitude of all low-energy features are true
predictions of the theory.

The spin resonance leads to a hump feature in $\Sigma''(\epsilon)$ for excitation energies between $20-30\, $meV at the $\alpha_1$-band, which is apparently observed in experiment \cite{Richard2008}. This feature is washed out in the $\alpha_2$-band. However the actual position of the resonance can not be obtained from such a vague energy range. 
In order to give a convincing quantitative statement, one has to consider the 
modification of the dispersion upon entering the superconducting state, which we
do now.

The values ${\bm k}_\epsilon $ determine an MDC-derived dispersion $\epsilon=\varepsilon_{\bm k}$, as can be seen as black curves in FIG.~\ref{fig2}(a) and (b).
The modification of the bare dispersion $\xi_{\bm k}$ to $\varepsilon_{\bm k}$
can be expressed in terms of the real part of the effective self-energy,
$\Sigma'$, that is defined by
$\varepsilon_{\bm k}=\xi_{\bm k} + \Sigma'(\varepsilon_{\bm k} ,{\bm k})$. 
The bare dispersion is not a measurable quantity. However,
as the bare dispersion is temperature independent, any {\it changes} of
$\Sigma'(\varepsilon_{\bm k} , {\bm k})\equiv \Sigma'_{\bm k}$ 
with temperature can be extracted experimentally by taking
the difference between two MDC-derived dispersions at fixed ${\bm k}$, provided again
the momentum dependence of the self energy is weak.
Comparing the normal and superconducting state dispersion, the influence of 
the resonance can then be quantified using e.g. the relation
\begin{equation}
\label{eqn13}
{\Sigma'_{\bm k}}^{\rm 15K} -{\Sigma'_{\bm k}}^{\rm 50 K}
=\varepsilon^{\rm 15K}_{{\bm k} }- \varepsilon^{\rm 50K}_{{\bm k} }.
\end{equation}
It is common to plot $\Sigma'(\epsilon)\equiv \Sigma'^T_{{\bm k}_\epsilon}
-\Sigma'^{T_{\rm ref}}_{{\bm k}_\epsilon}$ for a given reference temperature $T_{\rm ref}$ well inside the normal phase.

In FIG.~\ref{fig4} the MDC-derived dispersions of the $\alpha_{1,2}$ and the $\beta_1$ band in the normal and superconducting state are shown. The real part of the effective self energies in the respective inset feature a broad maximum at $\approx 50 \, $meV as well as a peak at lower energies. 
This peak 
appears due to coupling to the resonance at energy $\Delta^{\bm Q} + \Omega_{res}^{T}$, where $\Delta^{\bm Q}$ is the 
gap at the corresponding Fermi surfaces that are nested by an antiferromagnetic
wavevector. The shape as well as the absolute value of the real part of the effective self energy agree with experimental data \cite{Richard2008,Hasan2008}.\\

\noindent {\bf Tunnelling spectra}\\
\noindent
Another useful tool to obtain information on the density of states (DOS) is the scanning tunnelling spectroscopy (STS). 
Apart from the magnitude of the superconducting gap, also features due
to the interaction with spin excitations can be expected \cite{Eschrig2006}.
Since we calculate the spectral function in the entire Brillouin zone, 
the evaluation of tunnelling spectra of superconductor-insulator-normal metal (SIN) and 
superconductor-insulator-superconductor (SIS) tunnelling junctions  is straightforward. The SIN tunnelling current $I(V)$ is given by
\begin{eqnarray}
\label{eqn14}
I(V)=
\sum_{\bm k}
|M_{\bm k}|^2
\int\limits_{-\infty}^\infty  \frac{d\epsilon}{2\pi} A( \epsilon , \bm k) [f(\epsilon-eV)-f(\epsilon)].
\end{eqnarray}
For incoherent (momentum non-conserving) tunnelling we assume a constant $\left|M_{\bm k}\right|^2=\left|M_0\right|^2$.
The results for the differential SIN tunnelling conductance $\rho_{SIN}=dI/dV$ is shown in the top panel of FIG.~\ref{fig6} for varying coupling strength.
 \begin{figure}[t]
  \begin{center}
     \epsfig{figure=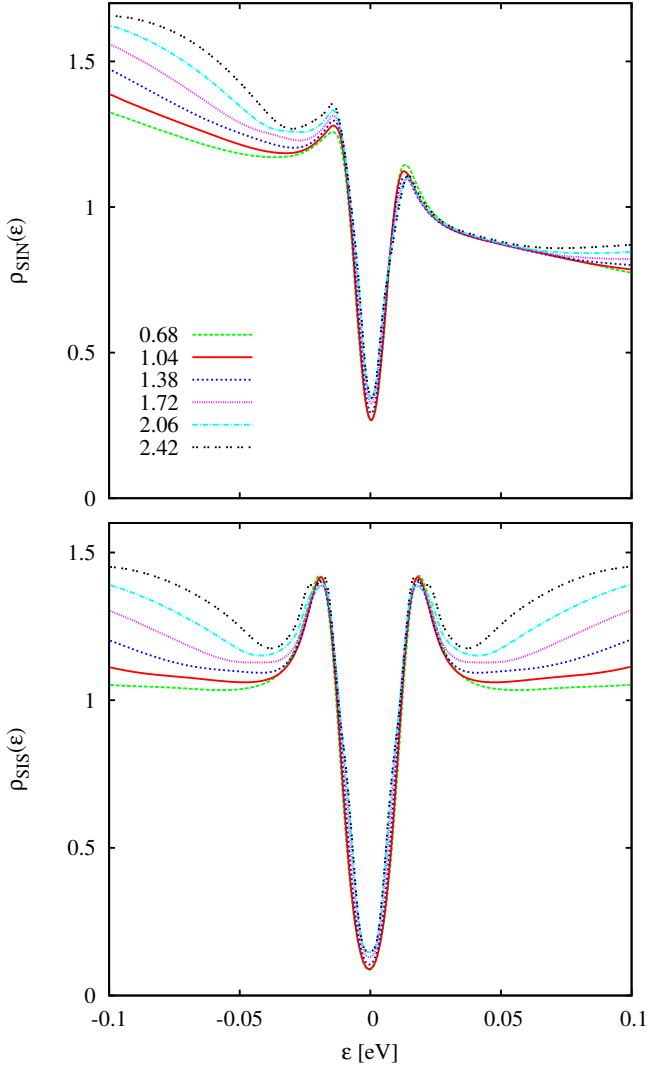, angle=270, width=0.5\textwidth}
    \caption{\textbf{STS tunnelling spectra}: Differential tunnelling conductance for different coupling constants 
$g^2\chi_0=0.68 ... 2.42 \times 10^3 \mu_{\rm B}^2 $eV~K
for the case of incoherent tunnelling at $T=15K$. Units are $e|M_0|^2$ for an SIN junction (top) and $e|T_0|^2$ for an SIS junction (bottom).}
    \label{fig6}
  \end{center}
\end{figure}
With increasing coupling constant a dip-hump structure develops and weight is shifted into the peak at the occupied side. Furthermore, the hump position moves towards the chemical potential. Analogously to the cuprates, coupling predominantly influences the occupied side \cite{Campuzano1999, Bogdanov2000}. 

For a SIS junction and the case of incoherent single particle tunnelling we
have with $A(\epsilon )=\sum_k A(\epsilon, \bm k)$
\begin{eqnarray}
\label{eqn14a}
I(V)= \left|T_0 \right|^2
\int\limits_{-\infty}^\infty  \frac{d\epsilon}{2\pi}
A(\epsilon ) A( \epsilon +eV ) [f(\epsilon)-f(\epsilon+eV)]
\end{eqnarray}
and again calculate the differential tunnelling conductance $dI/dV$, which is shown in the bottom panel of FIG.~\ref{fig6}. Due to the symmetry of Eq.~(\ref{eqn14a}), the dip-hump feature is now strong on both sides.

However, in both junctions we see no well pronounced dip for the coupling strength $g^2 \chi_0 = 1.17 \times 10^{3} \mu_{\rm B}^2$eV~K  which is the value we obtain from our comparison with ARPES experiments.\\

\noindent {\bf Discussion}\\
\noindent
From the knowledge of the experimental value of $\chi_0$ it is possible to extract the coupling constant $g$. This value is not known 
for Ba$_{1-x}$K$_x$Fe$_2$As$_2$}, however it has been determined
for optimally doped BaFe$_{1.85}$Co$_{0.15}$As$_2$ \cite{Inosov2009}. 
In order to obtain a rough estimate of the coupling constant $g$, we
insert this experimentally determined value $\chi_0=(3.8 \pm 1.0)\times 10^4 \,\mu_{\rm B}^2$eV$^{-1}$K and obtain $g\approx 0.15-0.2\, $eV.
We also have performed extended calculations of the temperature dependence of
the self energy effects discussed above, which are presented in the appendix.
From our results we conclude that the experimentally observed modifications of
the electronic dispersion when entering the superconducting state
can be explained within a model where a continuum of strongly incoherent
high-energy spin fluctuations provide the pairing interaction, whereas
the low-energy spin-fluctuation modes lead to 
the observed electronic dispersion features. The coupling strength that is required to 
explain the experimental data is large enough to allow for a magnetic pairing
scenario in iron-pnictide superconductors.

{\small
\subsection{Methods}
We employ a tight-binding fit in orbital basis that was obtained from the DFT band structure of BaFe$_2$As$_2$ by Ref.~\cite{Graser2010}. The largest contribution to the density of states comes from the five Fe-d-orbitals in the energy region we consider, which allows to reduce the Hamiltonian
\begin{equation}
H_0=\sum_{\bm k \sigma, mn} \,d_m^\dagger(\bm k \sigma ) \left[ \xi_{mn}(\bm k ) + \delta_{mn}\,\epsilon_n\right] \,d_n(\bm k \sigma), \label{eqn4}
\nonumber
\end{equation}
to this basis.
Here $d^\dagger_m(\bm k \sigma)$ creates an electron with momentum $\bm k$ and spin $\sigma $ [we denote $k =(\bm k, \sigma)$] in the orbital $m$,
where $m=1\ldots 5$ corresponds to the five orbitals
$d_{xz},\,d_{yz},\,d_{xy},\,d_{x^2-y^2},\,d_{3z^2-r^2}$.
The parameters $\xi_{mn}$ and $\epsilon_n$ are listed in Ref.~\cite{Graser2010}.
The chemical potential at zero temperature in the undoped state is at $\mu_{\rm c}=0$.
The canonical transformation
$d^\dagger_\mu(\bm k \sigma)=\sum_{m} a_\mu^m(\bm k) \, d^\dagger_m(\bm k \sigma)$,
where $a_\mu^m(\bm k) =\langle m{\bm k}|\mu {\bm k}\rangle $, diagonalises
the Hamiltonian, leading to eigenvalues $\xi_\mu(\bm k)$ and eigenvectors
$a_\mu^m(\bm k) $, where $\mu$ represents the band index.
We assume that the set of five eigenvectors for each $\bm k$
is orthonormal, $\sum_m a_\mu^m(\bm k)^\ast a_\nu^m (\bm k) =\delta_{\mu \nu}$, 
and use completeness, $\sum_\mu a_\mu^m(\bm k) a_\mu^n (\bm k)^\ast =\delta_{mn}$.
In order to simulate hole doping we apply a rigid shift of the chemical potential by an amount of $\delta \mu_{\rm c} =-50\, $meV. This leads to the appearance of additional hole pockets around $(\pi,\pi,k_z)$ that favour an $s^\pm$-state \cite{Graser2010}. 

The unperturbed Green's function is diagonal in band index, with normal
(diagonal)
and anomal (off-diagonal) components $G^{(0)}_\mu (\epsilon, k)$ and $F^{(0)}_\mu (\epsilon, k)$.
The renormalised Green's functions, $G_{\mu \nu}$ and
$F_{\mu \nu}$, are not diagonal in band index due to interband interactions introduced by spin fluctuations.
The Green's functions in an orbital basis, $G_{mn}$ and
$F_{mn}$, are related to those in a band representation, $G_{\mu \nu} $ and
$F_{\mu \nu}$, by
\begin{eqnarray}
\nonumber
G_{mn}(\epsilon, k) &=& \sum_{\mu,\nu} a_\mu^m(\bm k) a_\nu^n(\bm k)^\ast G_{\mu \nu}(\epsilon, k),\\
F_{mn}(\epsilon, k) &=& \sum_{\mu,\nu} a_\mu^m(\bm k) a_\nu^n(-\bm k) F_{\mu \nu}(\epsilon, k).
\nonumber
\end{eqnarray}
Note that due to $\xi_{mn}(-{\bm k}) = \xi_{mn}({\bm k})^\ast $ 
the eigenvectors can be chosen such that $a_\nu^n(-\bm k)=a_\nu^n(\bm k)^\ast $ holds.
 \\

Magnetic excitations can be described in terms of the dynamic spin susceptibility $\chi^{mq}_{np}$, 
where the measured susceptibility $\chi$ is the sum over all orbital contributions $\chi=\sum_{mn} \chi^{mm}_{nn}$ \cite {Maier2009}. Since the magnetic susceptibility is in particular enhanced for intra-orbital coupling \cite{Stanescu2008}, 
we assume that the elements $\chi_{nn}^{nn}$ of the susceptibility dominate, neglect in our model all inter-orbital contributions, i.e. $\chi\approx \sum_n \chi_{nn}^{nn}$, and take only those orbitals into account where the coupling is most pronounced ($n=xz,yz,xy$). 
The neglected contributions have a small spectral weight 
in the considered energy range $\omega <100 $meV, and are thus nearly unaffected by
superconductivity. Taking them into account leads to a simple energy and momentum
independent renormalisation of the low-energy electronic spectra (see below).
In order to determine the relative weight between $w_{xz}=w_{yz}$  and $w_{xy}$
we studied the behaviour of the electronic dispersion near the $X$ point of the
Brillouin zone. We find that experiments are best reproduced when the weight is
roughly equal. We thus performed our calculations for
$w_{xz}=w_{yz}=w_{xy}=w_T$ and $\chi_T^{xz}=\chi_T^{yz}=\chi_T^{xy}=\chi_T$. 
In order to determine the ratio between $\chi_T$ and $w_T$ we 
employ the total moment sum rule for the dynamical structure factor, 
as the temperature variation of the total moment is negligible.
With the above assumptions, we can
separate the sum-rule into its orbital parts. Employing the fluctuation dissipation
theorem that connects the 
dynamical spin structure factor with the susceptibility,
$S(\bm q, \omega)=2 \hbar \sum_n \, \chi_{sc,n}''(\bm q,\omega)/(1-e^{-\hbar\omega/k_B T})$, the ratio $w_T/\chi_T$ is determined by the equality 
\begin{eqnarray}
  \int_{-2\bar \Delta}^{2\bar \Delta} d\omega \int d^3\bm q \, S_{r,n}(\bm q, \omega)=\int_{-2\bar \Delta}^{2\bar \Delta}d\omega \int d^3\bm q\,S_{c,n}(\bm q, \omega)
\nonumber
\end{eqnarray}
with $S_{c(r),n}(\bm q, \omega)=2 \hbar \chi_{c(r),n}''(\bm q,\omega)/(1-e^{-\hbar\omega/k_B T})$ and a temperature dependent order parameter $\bar \Delta \propto \sqrt{1-T/T_c}$. \\

We couple electrons to the spin fluctuation spectrum with an energy and momentum independent coupling constant $g$ (instantaneous and local coupling). 
The retarded diagonal and off-diagonal self-energies are then given by 
\begin{equation}
\Sigma^R_{mn} =  \delta_{mn} \Sigma^R_n , \qquad
\Phi^R_{mn} =  \delta_{mn} \Phi^R_n ,
\nonumber
\end{equation}
written in terms of retarded ($R$) and Keldysh ($K$) Green's functions,
\begin{eqnarray}
\nonumber
\Sigma^R_{n}&=&-\frac{i}{2}g^2 \left(G^K_{nn} \ast \chi_n^R + G^R_{nn} \ast \chi_n^K\right),\quad\label{eqn9} \\
\Phi^R_{n}&=&-\frac{i}{2}g^2 \left(F^K_{nn} \ast \chi_n^R + F^R_{nn} \ast \chi_n^K\right),  \quad\label{eqn10}
\nonumber
\end{eqnarray}
with $(A\ast B) (\epsilon, \bm k)= \sum_{\omega,\bm q} A(\epsilon-\omega, \bm k - \bm q)\,B(\omega,\bm q)$ as explained in Ref.
\cite{Eschrig2006},
and with the diagonal susceptibility, $\chi_n\equiv \chi_{nn}^{nn}$.
In the normal and superconducting state the susceptibility is given by $\chi_n=\chi_c$ and $\chi_n=\chi_{sc}$ respectively,
where $n=1,2,3$. We calculate the convolutions 
numerically by fast Fourier transform, using bare Green's functions 
(a procedure supported by the numerical studies in Ref. \cite{Vilk1997})
with a broadening parameter $\delta = 4\, $meV.

Using the completeness relation $\sum_\mu a_\mu^m a_\mu^{n\ast} = \delta_{mn}$, 
we can write the Dyson equation for the
renormalised retarded Green's function in Nambu-Gor'kov space,
\begin{eqnarray}
{\hat G}^{R^{-1}}_{mn}(\epsilon,k)&=&
{\hat G}^{(0)R^{-1}}_{mn}-\hat \Sigma^R_{mn} 
=\sum_\mu a_\mu^m a_\mu^{n\ast} \left( \hat G_\mu^{(0)R^{-1}}-\hat \Sigma_n \right) \nonumber
\end{eqnarray}
in the following way
\begin{eqnarray}
&&{\hat G}^{R^{-1}}_{mn}(\epsilon,k)
=\sum_\mu a_\mu^m(\bm k) a_\mu^n(\bm k)^\ast \nonumber \\
&& \quad \times \left[(Z^{n}(\epsilon,\bm k) \, (\epsilon + \imath \delta) \hat {\mathbbm{1}} - \xi^{n}_\mu(\epsilon,\bm k) \hat \sigma_z - \Delta^{n}(\epsilon,\bm k) \hat \sigma_x \right] . \qquad \label{eqn5}
\nonumber
\end{eqnarray}
where $\hat {\mathbbm{1}}$ and $\hat \sigma_i$ are the 2$\times$2 unit matrix and the
Pauli matrices in Nambu space, respectively.
The renormalised dispersion and order parameter as well as the renormalisation function are given in terms of the retarded diagonal and off-diagonal self energies,
\begin{eqnarray}
\nonumber
 	\xi_\mu^{n}(\epsilon, \bm k)&=&\xi_\mu(\bm k)+\frac{\Sigma^R_{n}(\epsilon,\bm k)+\Sigma_{n}^{R}(-\epsilon,-\bm k)^\ast}{2}, \label{eqn6} \\
\nonumber
	\Delta^{n}(\epsilon,\bm k )&=&\Delta_{\bm k}+\Phi^R_{n}(\epsilon,\bm k),  \label{eqn7}\\
\nonumber
	Z^{n}(\epsilon,\bm k)&=&1-\frac{\Sigma^R_{n}(\epsilon,\bm k)-\Sigma_{n}^{R}(-\epsilon,-\bm k)^\ast }{2(\epsilon +\imath \delta)}, \label{eqn8}
\end{eqnarray}
using the fact that $\Phi^R_{n}(\epsilon,\bm k)=\Phi^{R}_{n}(-\epsilon,-\bm k)^\ast $. \\
The equation for ${\hat G}^{R^{-1}}_{mn}(\epsilon,k)$
is inverted numerically to obtain ${\hat G}^{R}_{mn}(\epsilon,k)$. 
The spectral function is then obtained from the retarded Green's function via
\begin{equation}
A(\epsilon,\bm k)=-\frac{1}{\pi} \mbox{Im} \left( \sum_m \left[\hat G^R_{mm}(\epsilon,\bm k) \right]_{11} \right). \label{eqn11}
\nonumber
\end{equation}
For the numerical calculations we use a $2^9\times2^9\times2^3$ k-mesh and $2^7$ points in energy space. 

A high-energy cutoff of $\omega_c= 200 \, \text{meV}$ was introduced in the
spectrum of spin excitations. The exact value of
this cutoff, as well as spectral weight beyond this cutoff and the precise variation of the spin susceptibility in the vicinity of the cutoff energy
are however not of importance, as
any change in the high-energy contributions only leads to an additional renormalisation factor $Z^{HE}$ for the low-energy electronic excitations. 
Thus as the high-energy part of the susceptibility varies with temperature, so does $Z^{HE}$. Taking the dispersion relation $\varepsilon_{\bm k}$ and the imaginary part of the effective self energy $\Sigma''_{\epsilon}(\bm k)$ as obtained from 
Eq.~(\ref{eqn12}) in the main text,
the spectral function in the superconducting state can be well approximated by 
\begin{eqnarray}
 	 A_{\epsilon}(\bm k)= -\frac{1}{\pi}  \mbox{Im} \left(\frac{1}{Z_\epsilon (\bm k) (\epsilon+\imath\delta) - \varepsilon_{\bm k}+\frac{\Delta_{\epsilon}(\bm k) ^2}{Z_\epsilon (\bm k) (\epsilon+\imath \delta) + \varepsilon_{\bm k}}}\right),
\nonumber
\end{eqnarray}
where $\Delta_{\epsilon}(\bm k)$ is the renormalised order parameter. 
The renormalised value of the gap is obtained 
by considering the quantity $\Delta_\epsilon (\bm k)/Z_\epsilon (\bm k )$.
The total renormalisation factor $Z_\epsilon(\bm k)$ can be expressed as a product of high- and low-energy renormalisations, $Z_\epsilon (\bm k)=Z^{HE}(\bm k)Z^{LE}_\epsilon(\bm k) $, with an energy-independent (on the low energy scale) $Z^{HE}(\bm k)$, and with
$Z^{LE}_\epsilon (\bm k) =1-\Sigma''_{\epsilon}(\bm k)/(\epsilon + \imath \delta)$. In order to compare the results for different temperatures we have to include the high energy contributions to the renormalisation of the electronic dispersions, i.e. $ \varepsilon_{\bm k}^{new}=\varepsilon_{\bm k}/Z^{HE}(\bm k)$. 
The high-energy renormalisation factor is only weakly momentum dependent due to
the weak momentum dependence of the susceptibility at high energies, and consequently we
neglect this momentum dependence in our calculation.
We are able to reproduce the experimentally observed high-energy part of the dispersions with 
$Z^{HE}(T=50K)=1$, and determine $Z^{HE}$ for the other temperatures so that the 
dispersions merge the normal state dispersion at $T=50K$ for high energies.
For the temperatures $T=15-35K$ $Z^{HE}$ varies slowly between $0.9-1$ as it is shown in Fig. \ref{S5}. Its value is $<1$ to account for the overestimate
of $\chi''$ near $\omega_c$ at lower temperatures.
\begin{figure}[hbtp]
  \begin{center}
	\epsfig{figure=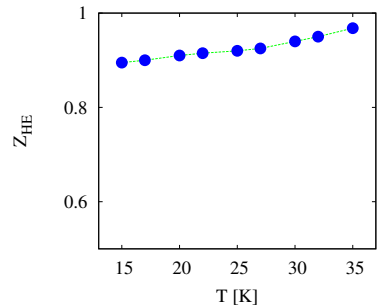, angle=270, width=0.3\textwidth}
    \caption{Temperature dependence of the high energy renormalisation factor $Z^{HE}$ corresponding to a high-energy cutoff for the spin excitation spectrum of $\omega_c=200$ meV.}
    \label{S5}
  \end{center}
\end{figure}

}

\centerline{\bf Acknowledgements}
We would like to thank Siegfried Graser for communications and clarifications with respect the parameter values in early versions of Ref.~\cite{Graser2010}.
We also would like to thank Jared Cole and Gerd Sch\"on for helpful discussions.

\appendix
\begin{widetext}
$\;$\\[20cm]
\noindent {\bf TEMPERATURE DEPENDENCE OF THE SPIN FLUCTUATION SPECTRUM AND THE RESONANCE MODE}\\
\noindent

We model the spin fluctuation spectrum by the procedure explained in the main article. The functional form of $\chi''(\omega,\bm q)$ is given by Eqs.~(1) and (2) in the paper. In FIG.~\ref{S1} the temperature dependence of the bosonic mode is shown. We present the energy dependence of the dynamic spin susceptibility at fixed momentum $\bm q=\bm Q$, where $\bm Q$ is the antiferromagnetic wave vector, as well as the energy-momentum dependence of the spectrum. With decreasing temperature, spectral weight of the two-particle excitation spectrum is shifted into the energy region $|\epsilon|<2\bar \Delta(T)$, leading to a well pronounced resonance feature that is sharp both in momentum and energy. Near $T_c$ the resonance disappears and the energy dependence obeys the same shape as the normal state dispersion at $T=50K$.  

\begin{figure}[hbtp]
  \begin{center}
	\includegraphics[width=\textwidth]{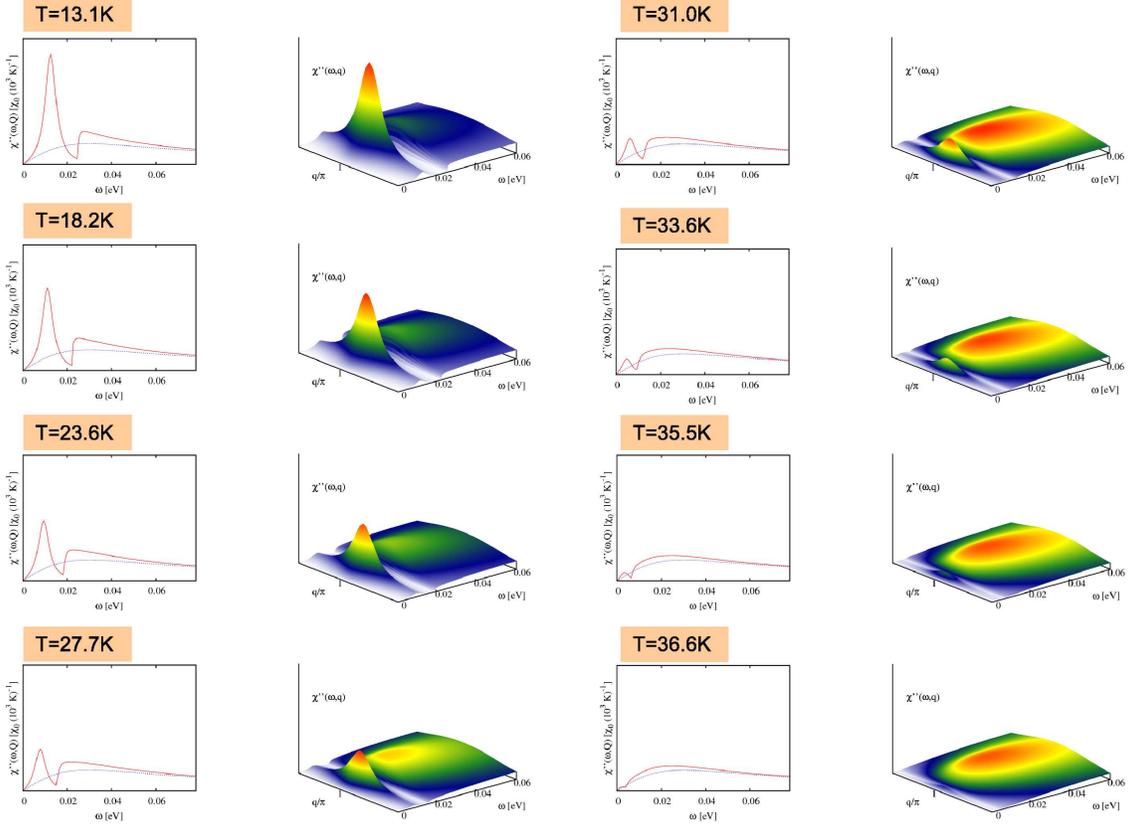}
    \caption{\textbf{Dependence of the spin fluctuation spectrum on temperature.} On the left of each panel the shape of the dynamic spin susceptibility at the antiferromagnetic wave vector, $\chi''(\omega,\bm Q)$, is shown (red line). The blue curve corresponds to the normal state spectral function at $T=50K$. On the right hand side the energy and momentum dependence around the wave vector $\bm Q$ is presented.}
    \label{S1}
  \end{center}
\end{figure}

\newpage
 \noindent {\bf INFLUENCE OF THE BOSONIC SPECTRAL SHAPE ON THE ELECTRONIC SPECTRA}\\
\noindent

In addition to the results in the paper, which are given for the temperatures $T=15K$ and $T=50K$, we performed the same calculations for various temperatures ($T=6.8-36.6K$).  The appearance of the bosonic resonance leads to an effect on the electronic dispersion which is characterised by the development of a peak in the real part of the effective self energy as well as a hump feature in the imaginary part of the effective self energy. Both effects clearly show the same temperature dependence as the resonance in the bosonic spectrum as presented in FIG~\ref{S2} and FIG~\ref{S3}.\\ 

\begin{figure}[hbtp]
  \begin{center}
	\epsfig{figure=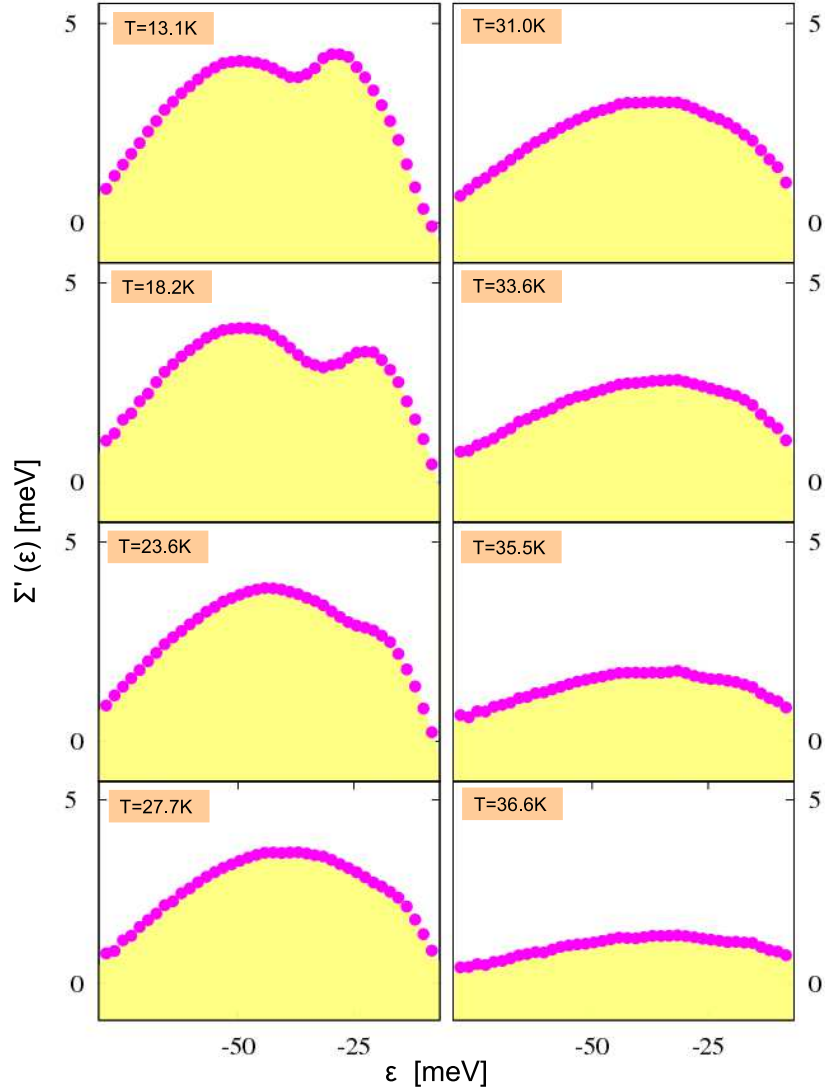, angle=0, width=0.7\textwidth}
    \caption{\textbf{Temperature dependence of the real part of the effective self energy} at the $\beta_1$-band. With increasing temperature, the self energy shows the development of a broad hump and a peak appearing at low energies.}
	\label{S2}
  \end{center}
\end{figure}

\begin{figure}[hbtp]
  \begin{center}
	\epsfig{figure=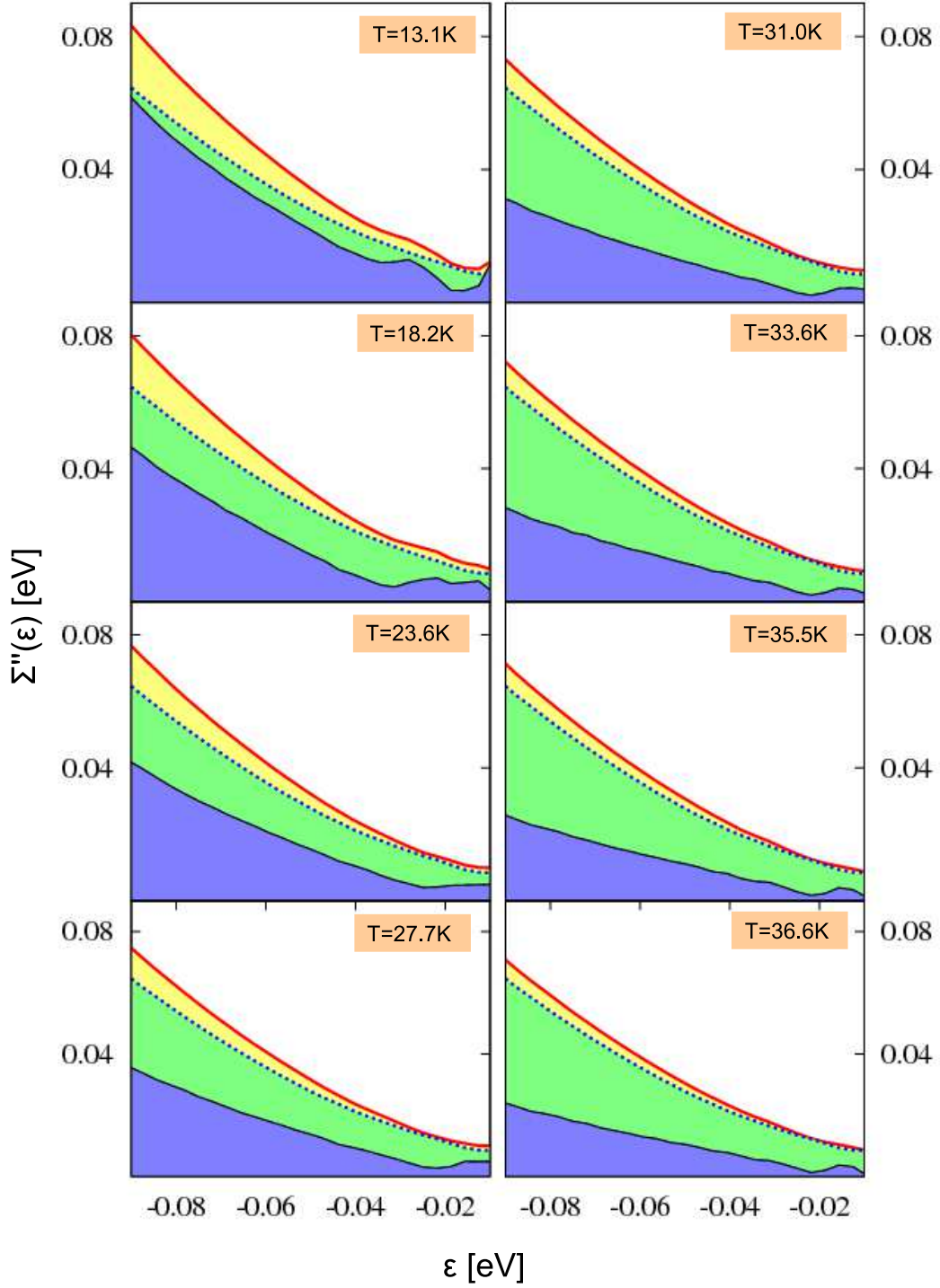, angle=0, width=0.9\textwidth}
	 \caption{\textbf{Temperature dependence of the imaginary part of the effective self energy} at the $\alpha_1$-band. We present the results for each temperature (red line) compared to the normal state imaginary part of the effective self energy at $T=50K$ (blue line). The black curve presents the difference of both, scaled by a factor of 3.}
   \label{S3}
  \end{center}
\end{figure}

\end{widetext}

\end{document}